\documentclass[%
 aip,
rsi,%
 amsmath,amssymb,
preprint,
]{revtex4-1}

\usepackage{graphicx}
\usepackage{dcolumn}
\usepackage{bm}
\usepackage{siunitx}
\usepackage{braket}
\usepackage{fancyhdr}
\newcommand{\angstrom}{\mbox{\normalfont\AA}}

\begin{document}


\title[Calculating the Band Structure of 3C-SiC Using sp\textsuperscript{3}d\textsuperscript{5}s\textsuperscript{*}+$\Delta$ Model]{Calculating the Band Structure of 3C-SiC Using sp\textsuperscript{3}d\textsuperscript{5}s\textsuperscript{*}+$\Delta$ Model}

\author{Murat Onen}
\thanks{These authors contributed equally to this work. Authors to whom correspondence should be addressed: monen@mit.edu and turchett@mit.edu}
\author{Marco Turchetti}%
\thanks{These authors contributed equally to this work. Authors to whom correspondence should be addressed: monen@mit.edu and turchett@mit.edu}

\affiliation{\textsuperscript{1}Research Laboratory of Electronics, Massachusetts Institute of Technology, Cambridge, MA 02139, USA}%

\date{\today}

\begin{abstract}
We report on a semi-empirical tight binding model for 3C-SiC including the effect of sp\textsuperscript{3}d\textsuperscript{5}s\textsuperscript{*} orbitals and spin orbital coupling. In this work, we illustrate in detail the method to develop such a model for semiconductors with zincblende structure, based on Slater-Koster integrals, and we explain the optimization method used to fit the experimental results with such a model. This method shows high accuracy for the evaluation of 3C-SiC band diagram both in terms of the experimental energy levels at high symmetry points and the effective masses.

\end{abstract}

\maketitle

\section{\label{sec:level1}Introduction}

Silicon Carbide (SiC) exhibits strong chemical bonding and physical stability. 3C polytype of SiC is of particular interest due to its superior electronic properties such as high electron mobility and saturation velocity which makes it a perfect candidate for building devices that have to withstand harsh environments.\cite{Harris,Shur,Fan} In particular, it is widely used in high voltage and high temperature semiconductor industries, astronomy and, as it is resistant to radiation, in nuclear reactors.\cite{Saddow} Therefore, the understanding of its electronic structure is critical for the improvement of existing SiC based technologies and the development of new applications.

Several models have been previously used to fit the experimental data to reconstruct SiC band diagram, including Density Functional Theory (DFT) with local density approximation (LDA) \cite{Kackell,Theodorou}, Hartree-Fock-Slater model using discrete variational method \cite{Lubinsky} and empirical pseudopotential method \cite{Junginger,Hemstreet}. This work implements semi-empirical tight binding model (SETBM) for fitting experimental data to calculate the band structure of semiconductors, which proved its reliability through the years. 

SETBM approach has previously been applied for SiC in literature, but was limited to only including $sp^3$ orbitals entailing an 8$\,\times\,$8 Hamiltonian matrix \cite{Theodorou}. To improve the capacity of the model, an excited orbital $s^*$ has been included as well \cite{Vogl}. However, inclusion of the $d$ orbitals and the spin orbital coupling ($\Delta$) is required to portray the electron band structure in a more complete way that accounts for the splitting of the bands due to the lifted degeneracy with the spin-orbital coupling (SOC) effects. In this work, we propose an $sp^3d^5s^*+\Delta$ model for 3C-SiC based on Slater-Koster integrals\cite{Slater}. Similar models have been applied for III-V semiconductors by Jancu \cite{Jancu}. We first show how to construct the resulting 40$\,\times\,40$ Hamiltonian matrix and then how to optimize it using the experimental data.

\section{Semi-Empirical Tight-Binding Method (SETBM) for Zincblende Structures}

To evaluate the band diagram of 3C-SiC we build the Hamiltonian matrix using the linear combination of atomic orbitals (LCAO) model, considering only the nearest neighbor interactions which collapse the overlap matrix to identity.

\subsection{sp\textsuperscript{3}d\textsuperscript{5}s\textsuperscript{*} Model}

The tight binding Hamiltonian matrix is built by evaluating each interaction integral $H_{jk}= \braket{\Phi_j|\hat{H}|\Phi_k}$  between the nearest neighbor orbitals. In this notation  $\ket{\Phi_i}$ are the s,p,d and s* orbitals, and $\hat{H}$ is the full crystal interaction Hamiltonian. Using 10 orbitals (s, $p_x$, $p_y$,  $p_z$, $d_x$, $d_y$, $d_z$, $d_{x^2-y^2}$, $d_{3r^2-z^2}$) for 2 atoms (cation: $Si^+$ and anion $C^-$ ) result in a Hamiltonian matrix of $20\,\times\,20$ before the inclusion of SOC.

To evaluate these integrals we adopted the Slater-Koster notation \cite{Slater}, using $l=m=n=-\frac{1}{\sqrt{3}}$ as 3C-SiC has a zincblende structure. Fig. \ref{fig1} shows the full resulting 20$\,\times\,$20 Hamiltonian. Matrix entries are calculated as following: 

\begin{equation}
\begin{array}{lcl}
    V_{ss}=V_{ss\sigma} \\
    V_{sp}=-\frac{1}{\sqrt{3}}V_{sp\sigma}  \\
    V_{xx}=\frac{1}{3}V_{pp\sigma}+\frac{2}{3}V_{pp\pi} \\
    V_{xy}=\frac{1}{3}V_{pp\sigma}-\frac{1}{3}V_{pp\pi} \\
    V_{sd}=\frac{1}{\sqrt{3}}V_{sd\sigma}  \\
    V_{pd}=\frac{1}{3} \left[ V_{pd\sigma} -\frac{2}{\sqrt{3}}V_{pd\pi} \right] \\
    U_{pd}=\frac{1}{3} \left[ V_{pd\sigma} +\frac{1}{\sqrt{3}}V_{pd\pi} \right] \\
    W_{pd}=\frac{2}{3} V_{pd\pi} \\
    V_{dd}=\frac{1}{9} \left[ 3 V_{dd\sigma} +2V_{dd\pi} + 4V_{dd\delta} \right] \\
    \tilde{V_{dd}}=\frac{1}{3} \left[ 2V_{dd\pi} +V_{dd\delta} \right] \\
    U_{dd}=\frac{1}{9} \left[ 3 V_{dd\sigma} -V_{dd\pi} - 2V_{dd\delta} \right] \\
    W_{dd}=\frac{2}{3\sqrt{3}} \left[ -V_{dd\pi} + V_{dd\delta} \right] \\
    \end{array}
\end{equation}

The $\alpha$ and $\beta$ superscripts define whether it refers to an anion-to-cation integral or a cation-to-anion one. The s* integrals use the same notation of the s ones. Diagonal elements ($E_{ii}$) are the self integrals of the orbitals. The $g_i$ are the phase factor that take into account the fact that we are evaluating integrals respect to each nearest neighbor, that are defined as:

\begin{equation}
\begin{array}{lcl}
    g_{0}= 1+e^{-ikR_1}+e^{-ikR_2}+e^{-ikR_3} \\
    g_{1}= 1+e^{-ikR_1}-e^{-ikR_2}-e^{-ikR_3} \\
    g_{2}= 1-e^{-ikR_1}+e^{-ikR_2}-e^{-ikR_3} \\
    g_{3}= 1-e^{-ikR_1}-e^{-ikR_2}+e^{-ikR_3} \\
    \end{array}
\end{equation}

where $R_{1}= \left[ -\frac{a}{2} , -\frac{a}{2} , 0 \right]$ , $R_{2}= \left[ 0 , -\frac{a}{2} , -\frac{a}{2} \right]$ and $R_{3}= \left[ -\frac{a}{2} , 0 , -\frac{a}{2} \right]$, and $a$ is the lattice constant.

\subsection{Spin-Orbital Coupling}

In this model we also take into account of the spin-orbital coupling between p orbitals as explained by Datta \cite{Datta}. Spin-orbit interaction is responsible for lifting the degeneracy of the valence band and in the evaluation of the optical properties of the material. \cite{Theodorou} In this work, we considered only the contribution of p valence states since the one of excited d states is much smaller.\cite{Jancu} 

The introduction of spin-orbit interaction in the model is implemented distinguishing between $\uparrow$ and $\downarrow$ electrons and creating a matrix twice the rank with the introduction of two coupling parameters $\delta_a$ and $\delta_c$ (where $\delta_{a,c} = \Delta_{a,c}/3$). Such a $40\,\times\,40$ matrix $H$ can be defined starting from a matrix having two of the $H_{sp^3d^5s^*}$$20,\times\,20$ matrix portrayed in Fig.\ref{fig1} as diagonal elements and adding to it a coupling matrix $H^\Delta$:

\begin{equation}
H=
    \begin{bmatrix}
    H_{sp^3d^5s^*}       & 0 \\
    0       & H_{sp^3d^5s^*} 
\end{bmatrix}
+
    \begin{bmatrix}
    H_{11}^\Delta       & H_{12}^\Delta \\
    H_{21}^\Delta       & H_{22}^\Delta 
\end{bmatrix}
\end{equation}

where $H_{ii}^\Delta$ are defined in Fig.\ref{fig2}.

\subsection{Verification of the Model}

The model involving sp\textsuperscript{3}d\textsuperscript{5}s\textsuperscript{*}+$\Delta$ parameters has previously been applied to a set of semiconductors in Ref. \cite{Jancu}. As a verification of the model construction and optimization procedure of the coupling parameters, here we demonstrate the results obtained for GaAs. Table \ref{table1} shows that the coupling parameters and the accuracy of our model is comparable to that of reported in Ref. \cite{Jancu}.

\begin{table}[h!]
\begin{tabular}{c|c|c|c}
\hline
Parameter                 &Previous Work\cite{Jancu}              & This Work                         & Experimental\cite{Theodorou}    \\ \hline
$\Gamma_{6\textit{v}}$    & $\SI{-12.910}{\electronvolt}$         & $\SI{-13.070}{\electronvolt}$      & $\SI{-13.1}{\electronvolt}$    \\
$-\Delta_0$               & $\SI{ -0.340}{\electronvolt}$         & $\SI{ -0.339}{\electronvolt}$      & $\SI{ -0.341}{\electronvolt}$   \\
$\Gamma_{6\textit{c}}$    & $\SI{  1.519}{\electronvolt}$         & $\SI{  1.519}{\electronvolt}$      & $\SI{  1.519}{\electronvolt}$    \\
$\Gamma_{7\textit{c}}$    & $\SI{  4.500}{\electronvolt}$         & $\SI{  4.497}{\electronvolt}$      & $\SI{  4.53}{\electronvolt}$     \\
$\Gamma_{8\textit{c}}$    & $\SI{  4.716}{\electronvolt}$         & $\SI{  4.764}{\electronvolt}$      & $\SI{  4.716}{\electronvolt}$    \\
$X_{6\textit{v}}$         & $\SI{ -3.109}{\electronvolt}$         & $\SI{ -2.904}{\electronvolt}$      & $\SI{ -2.88}{\electronvolt}$    \\
$X_{7\textit{v}}$         & $\SI{ -2.929}{\electronvolt}$         & $\SI{ -2.790}{\electronvolt}$      & $\SI{ -2.80}{\electronvolt}$    \\
$X_{6\textit{c}}$         & $\SI{  1.989}{\electronvolt}$         & $\SI{  2.009}{\electronvolt}$      & $\SI{  1.98}{\electronvolt}$     \\
$X_{7\textit{c}}$         & $\SI{  2.328}{\electronvolt}$         & $\SI{  2.385}{\electronvolt}$      & $\SI{  2.35}{\electronvolt}$     \\
$L_{6\textit{v}}$         & $\SI{ -1.330}{\electronvolt}$         & $\SI{ -1.427}{\electronvolt}$      & $\SI{ -1.42}{\electronvolt}$    \\
$L_{4,5\textit{v}}$       & $\SI{ -1.084}{\electronvolt}$         & $\SI{ -1.180}{\electronvolt}$      & $\SI{ -1.20}{\electronvolt}$    \\
$L_{6\textit{c}}$         & $\SI{  1.837}{\electronvolt}$         & $\SI{  1.829}{\electronvolt}$      & $\SI{  1.85}{\electronvolt}$     \\
$L_{7\textit{c}}$         & $\SI{  5.047}{\electronvolt}$         & $\SI{  5.303}{\electronvolt}$      & $\SI{  5.47}{\electronvolt}$     \\ 
$m(\Gamma_{6c})$          & 0.067$m_0$                            & 0.067$m_0$                         & 0.067$m_0$                     \\ \hline
Avg. Accuracy                  & 97.11\%                               & 99.45\%                            & 100\%                          \\
\end{tabular}
\caption{Comparison between experimental values of the GaAs high symmetry points and effective masses, the corresponding values evaluated by Jancu et al. in Ref. \cite{Jancu} using SETBM and the same value evaluated using our model.}
\label{table1}
\end{table}

\section{Optimization for S\MakeTextLowercase{i}C}

Following the construction of the matrix, an optimization procedure is necessary to find the SETBM parameters for any given material. Here we provide a method that can be generalized to other materials as well. The optimization requires experimental data on energy levels at high symmetry points and effective masses in certain directions, a well-designed cost function and finally a physically meaningful initial point. As described in Ref. \cite{Jancu}, a good candidate for the initial point is the free-electron model generated parameters for the coupling energies. We have also observed that using other zincblende structures' coupling parameters as initial points produced good results.

We have built our cost function to take into account of both energy levels and effective masses at the same time. Since the cost comprises of points and curvatures to fit, this ensures a physically meaningful band diagram when initiated from the points described above. Following the definition of this cost function we perform constrained non-linear optimization using Nelder-Mead simplex algorithm \cite{Lagarias}. Other optimization methods such as genetic algorithms are found to be inefficient for this approach since the problem structure is sufficiently bounded and thus does not need high level of exploration.

\section{Results}

\begin{figure}[h!]
    \centering
    \includegraphics[width = \linewidth]{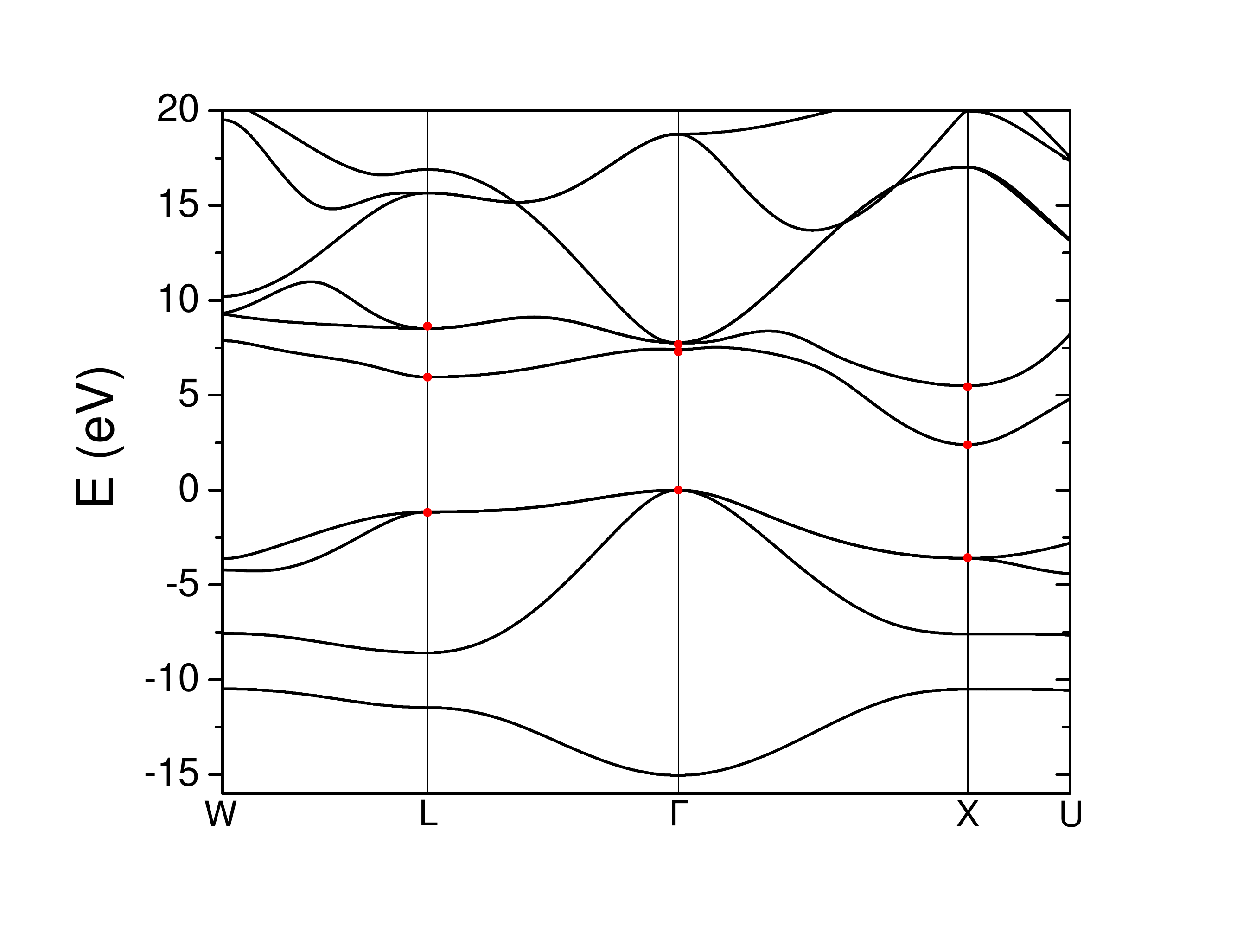}
    \caption{Electronic band structure for SiC calculated with sp\textsuperscript{3}d\textsuperscript{5}s\textsuperscript{*}+$\Delta$ model. Experimental values reported in Ref.\cite{Theodorou} are noted with red dots.}
    \label{fig:Band_Structure}
\end{figure}

Using the model described above, here we present the electronic band structure and the corresponding density of states calculated for 3C-SiC. Table \ref{table2} shows a comparison between the experimental values of high symmetry points and effective masses and the same values calculated using our model and a previous SETBM implementation from Theodorou et al. \cite{Theodorou}. It can be seen from Fig. \ref{fig:Band_Structure} that the conduction band minimum is at X point giving and indirect band structure with bandgap of 2.39 eV. Optimized model predicts the experimental results with a reasonably high average accuracy of 99.91\%.  Other than the high symmetry points, it can also be noticed that the band diagram shows a constant gap in the $\Lambda$ direction between valence and conduction bands as expected from reflectivity measurements.\cite{Madelung} Coupling parameters optimized for 3C-SiC can be found in Table \ref{table3}.

Specifically, for the band diagram of 3C-SiC the spin-orbital coupling does not play a major role as the splitting of the valence bands is 10.3 meV. The effect becomes more important when other physical parameters, such as dielectric function \cite{Theodorou}, are of interest. Mainly, not to lose the generality of the model, this effect is included in all of the calculations, obtaining small spin-orbital coupling parameters ($\delta_a$, $\delta_c$) as expected.

\begin{figure}[h!]
    \centering
    \includegraphics[width = \linewidth]{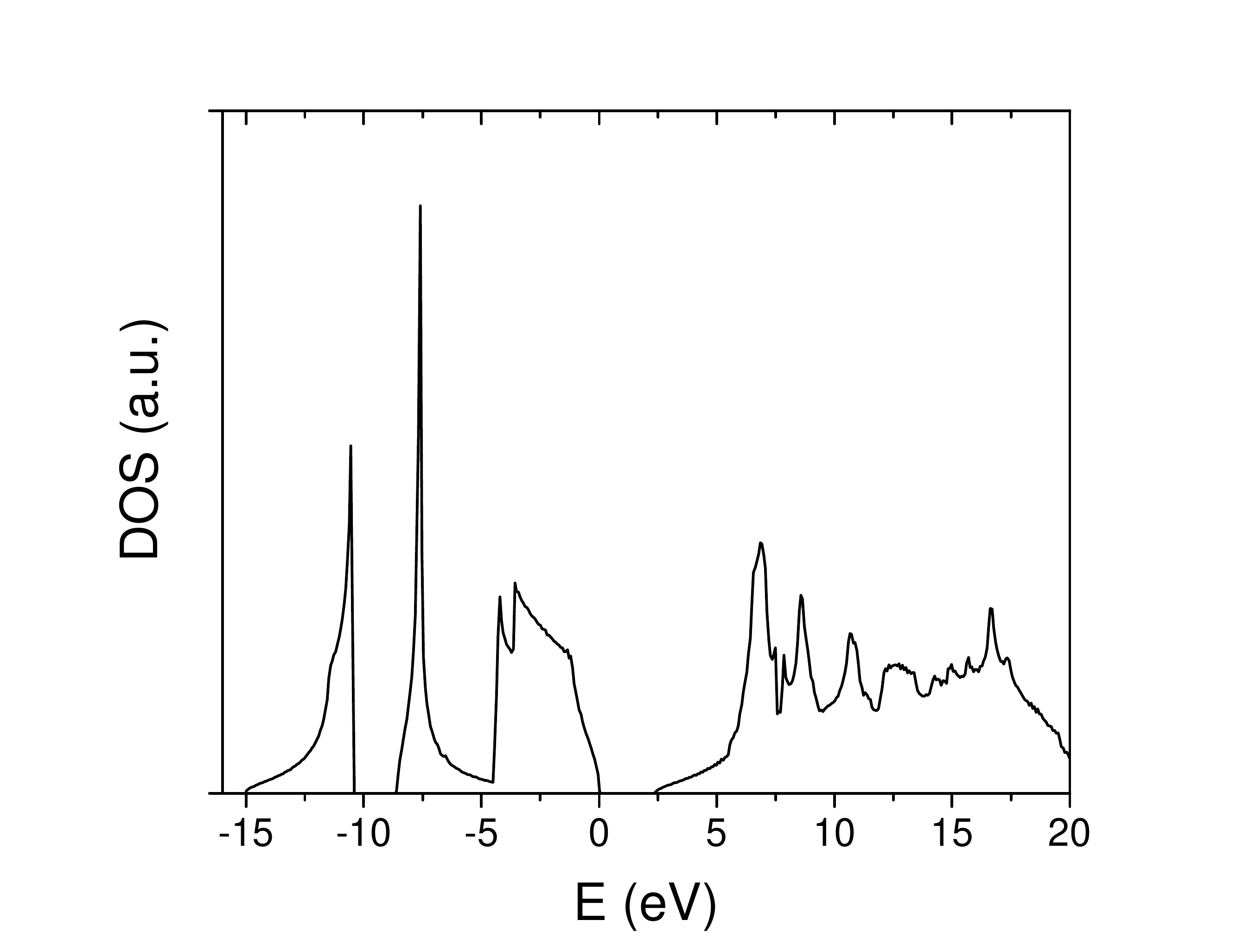}
    \caption{Density of states of SiC calculated with sp\textsuperscript{3}d\textsuperscript{5}s\textsuperscript{*}+$\Delta$ model described in this work.}
    \label{fig:DoS}
\end{figure}

\newpage
\begin{table}[h!]
\begin{tabular}{c|c|c|c}
\hline \hline
Parameter               & Previous Work\cite{Theodorou}       & This Work                              & Experimental\cite{Theodorou} \\ \hline
$\Delta_0$              & -                                   & $\SI{-10.3600}{\milli\electronvolt}$   & $\SI{-10.36}{\milli\electronvolt}$\\ 
$\Gamma_{1c} $          & $\SI{7.07}{\electronvolt}$          & $\SI{7.4000}{\electronvolt}$           & $\SI{7.40}{\electronvolt}$       \\
$\Gamma_{15c}$          & $\SI{8.98}{\electronvolt}$          & $\SI{7.7253}{\electronvolt}$           & $\SI{7.75}{\electronvolt}$      \\
$X_{5v}$                & $\SI{-3.2}{\electronvolt}$          & $\SI{-3.5977}{\electronvolt}$          & $\SI{-3.60}{\electronvolt}$    \\
$X_{1c}$                & $\SI{2.47}{\electronvolt}$          & $\SI{2.3900}{\electronvolt}$           & $\SI{2.39}{\electronvolt}$    \\
$X_{3c}$                & $\SI{5.64}{\electronvolt}$          & $\SI{5.5000}{\electronvolt}$           & $\SI{5.50}{\electronvolt}$    \\
$L_{3v} $               & $\SI{-1.18}{\electronvolt}$         & $\SI{-1.1522}{\electronvolt}$          & $\SI{-1.16}{\electronvolt}$   \\
$L_{1c}$                & $\SI{6.22}{\electronvolt}$          & $\SI{5.9400}{\electronvolt}$           & $\SI{5.94}{\electronvolt}$    \\
$L_{3c}$                & $\SI{9.11}{\electronvolt}$          & $\SI{8.5004}{\electronvolt}$           & $\SI{8.50}{\electronvolt}$   \\ 
$m_\parallel$           & -                                   & 0.6700$m_0$                            & 0.67$m_0$                  \\ 
$m_\bot$                & -                                   & 0.2500$m_0$                            & 0.25$m_0$                  \\ \hline 
Avg. Accuracy           & 93.6\%                              & 99.91\%                                & 100\%   \\
\hline \hline
\end{tabular}
\caption{Comparison between experimental values of the SiC high symmetry points and effective masses, the corresponding values evaluated by Theodorou et al. in Ref. \cite{Theodorou} using SETBM and the same value evaluated using our model.}
\label{table2}
\end{table}

\section{Conclusion}

We have reported the full implementation of a semi-empirical tight binding model for zincblende 3C-SiC with $sp\textsuperscript{3}d\textsuperscript{5}s\textsuperscript{*}$ orbitals and spin-orbit coupling ($\Delta$). Model parameters, initialized using free electron model parameters and optimized using experimental energy values and effective masses has shown 99.91\% average accuracy. Construction of the matrix and the optimization procedure can be further applied to other materials, in describing their electronic properties.

\begin{table}[h!]
\label{table3}

\begin{tabular}{c|c|c}
\hline \hline
Parameter                                   & GaAs               & SiC       \\ \hline
$\textit{a}$                                & 5.6532$\angstrom$  & 4.3596  $\angstrom$ \\ \hline
$\textit{E}_\textit{$s_\textit{a}$}$        & -6.0533            & -0.8766   \\
$\textit{E}_\textit{$s_\textit{c}$}$        & -0.3340            & -0.3421   \\
$\textit{E}_\textit{$p_\textit{a}$}$        &  3.3063            & 0.5489    \\
$\textit{E}_\textit{$p_\textit{c}$}$        &  6.2866            & 5.4488    \\
$\textit{E}_\textit{$d_\textit{a}$}$        & 13.3395            & 22.9218   \\
$\textit{E}_\textit{$d_\textit{c}$}$        & 13.3327            & 14.7797   \\
$\textit{E}_\textit{$s^*_\textit{a}$}$      & 19.3982            & 21.4411   \\
$\textit{E}_\textit{$s^*_\textit{c}$}$      & 19.3982            & 24.3075   \\  \hline
$\textit{V}_\textit{ss$\sigma$}$            & -1.7167            & -1.9875   \\
$\textit{V}_\textit{$s^*s^*\sigma$}$        & -3.9205            & -1.5826   \\
$\textit{V}_\textit{$s^*_as_c\sigma$}$      & -2.1479            & -6.9155   \\
$\textit{V}_\textit{$s_as^*_c\sigma$}$      & -1.3658            & -0.7085   \\ \hline
$\textit{V}_\textit{$s_ap_c\sigma$}$        & 2.6999             & 5.6044    \\ 
$\textit{V}_\textit{$s_cp_a\sigma$}$        & 2.9036             & 4.6564    \\ 
$\textit{V}_\textit{$s^*_ap_c\sigma$}$      & 2.2556             & 6.5528    \\ 
$\textit{V}_\textit{$s^*_cp_a\sigma$}$      & 2.5823             & 5.0141    \\ \hline
$\textit{V}_\textit{$s_ad_c\sigma$}$        & -2.7144            & -6.5282   \\ 
$\textit{V}_\textit{$s_cd_a\sigma$}$        & -2.4623            & -4.3586   \\ 
$\textit{V}_\textit{$s^*_ad_c\sigma$}$      & -0.6651            & -0.2985   \\ 
$\textit{V}_\textit{$s^*_cd_a\sigma$}$      & -0.5404            & -0.3126   \\ \hline
$\textit{V}_\textit{$pp\sigma$}$            & 4.3807             & 6.9700    \\ 
$\textit{V}_\textit{$pp\pi$}$               & -1.3874            & -2.2015   \\ \hline
$\textit{V}_\textit{$p_ad_c\sigma$}$        & -1.3147            & -3.9538   \\ 
$\textit{V}_\textit{$p_cd_a\sigma$}$        & -1.5263            & -6.0081   \\ 
$\textit{V}_\textit{$p_ad_c\pi$}$           & 2.1184             & 1.4686    \\ 
$\textit{V}_\textit{$p_cd_a\pi$}$           & 2.4926             & 3.1505    \\ \hline
$\textit{V}_\textit{$dd\sigma$}$            & -0.7282            & -1.1553   \\ 
$\textit{V}_\textit{$dd\pi$}$               & 1.6289             & 4.4417    \\
$\textit{V}_\textit{$dd\delta$}$            & -1.8121            & -4.9623   \\ \hline
$\Delta_a/3$                                & 0.1630             & 0.0071    \\ 
$\Delta_c/3$                                & 0.0823             & 0.0030    \\ 
\hline \hline
\end{tabular}
\caption{sp\textsuperscript{3}d\textsuperscript{5}s\textsuperscript{*}+$\Delta$ parameters used in our model for GaAs and SiC.}
\end{table}

\newpage
\newpage

\begin{acknowledgments}
The authors would like to acknowledge Prof. Qing Hu from Massachusetts Institute of Technology for the helpful discussions and guidance.
\end{acknowledgments}

\nocite{*}
\newpage
\bibliography{aipsamp}

\begin{figure*}
\centering
\includegraphics[width=0.61\pdfpagewidth]{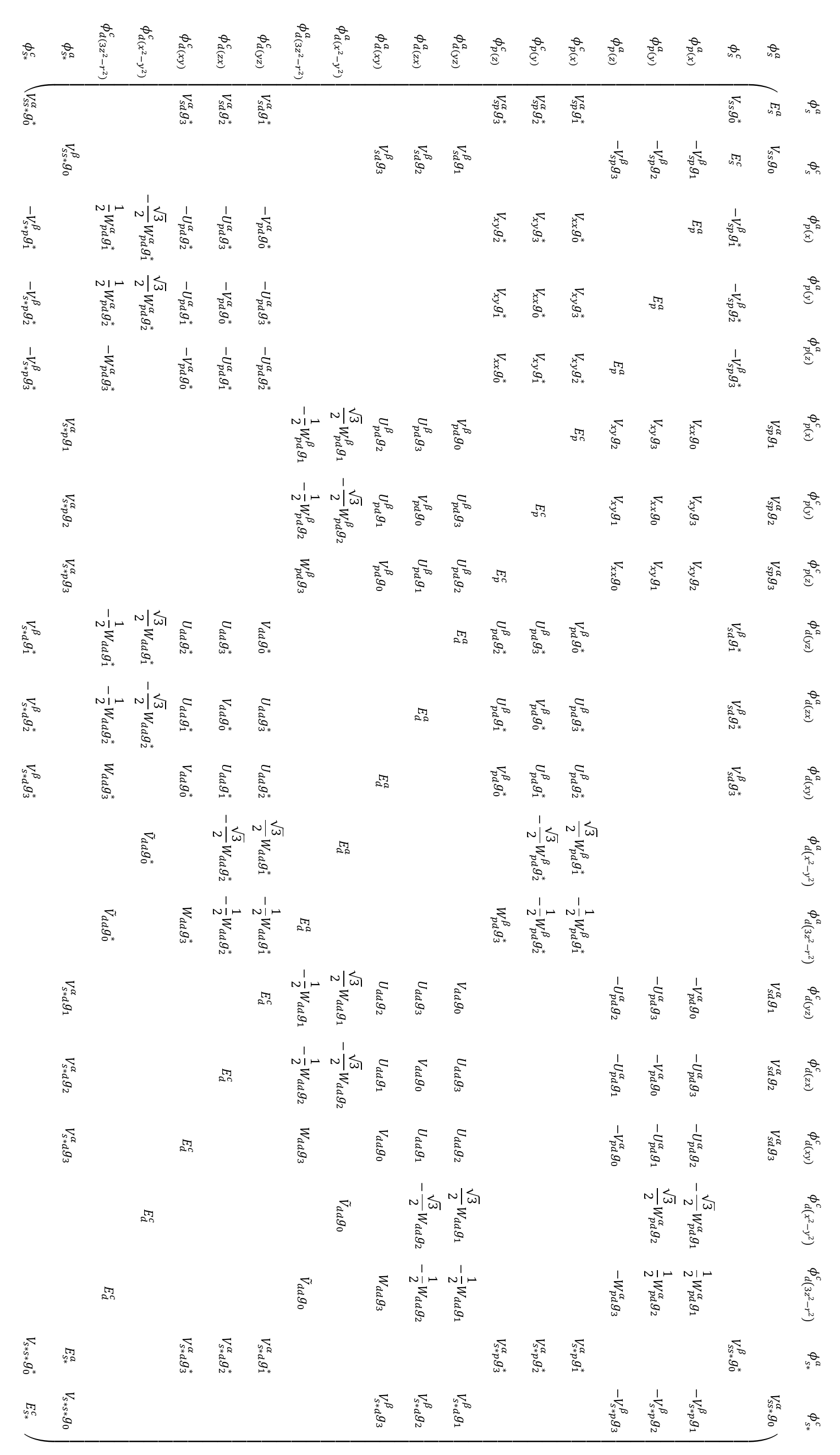}
\caption{\label{fig1} The 20x20 $H_{sp^3d^5s^*}$ matrix.}
\end{figure*}

\begin{figure*}
\centering
\includegraphics[width=0.64\pdfpagewidth]{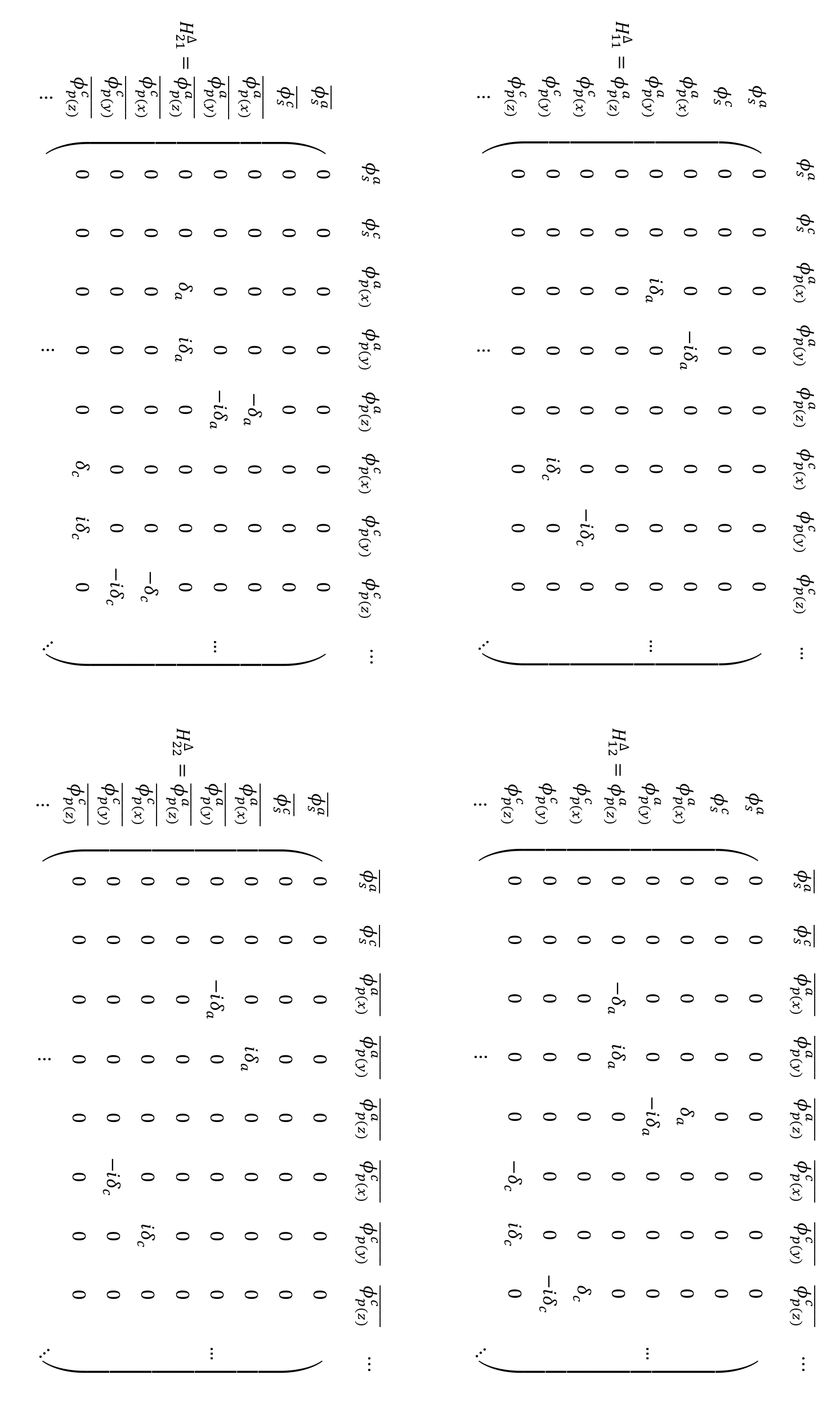}
\caption{\label{fig2} The spin-orbital matrix.}
\end{figure*}

\end{document}